Primitive Chain Network Simulations of Entangled Melts of Symmetric and Asymmetric Star Polymers in Uniaxial Elongational Flows

[1]*Yuichi Masubuchi, [2]Giovanni Ianniruberto, and [2]Giuseppe Marrucci

[1]Department of Materials Physics, Nagoya University, Nagoya 4648603, Japan
[2]Dipartimento di Ingegneria Chimica, dei Materiali e della Produzione Industriale, Università degli Studi di Napoli "Federico II", Piazzale Tecchio 80-80125 Napoli, Italy

*To whom correspondence should be addressed: mas@mp.pse.nagoya-u.ac.jp

ABSTRACT

Ianniruberto and Marrucci [Macromolecules, 46, 267-275, 2013] developed a theory whereby entangled branched polymers behave like linear ones in fast elongational flows. In order to test such prediction, Huang et al [Macromolecules, 49, 6694-6699, 2016] performed elongational measurements on star polymer melts, indeed revealing that, in fast flows, the elongational viscosity is insensitive to the molecular structure, provided the molecular weight of the "backbone" (spanning the largest end-to-end distance of the molecule) is the same. Inspired by these studies, we here report on results obtained with multi-chain sliplink simulations for symmetric and asymmetric star polymer melts, as well as calculations of the Rouse time of the examined branched structures (not previously determined by Ianniruberto and Marrucci). The simulations semi-quantitatively reproduce the experimental data if the Kuhn-segment orientation-induced reduction of friction (SORF) is accounted for. The observed insensitivity of the nonlinear elongational viscosity to the molecular structure for the same span molar mass may be due to several factors. In the symmetric case, the calculated Rouse time of the star marginally differs from that of the linear molecule, so that possible differences in the observed stress fall within the experimental uncertainty. Secondly, it is possible that flow-induced formation of hooked star pairs (or even larger aggregates) makes the effective Rouse time of the aggregate even closer to that of the linear polymer because the friction center moves towards the branchpoint of the star molecule. In the asymmetric case, it is shown that the stress contributed by the short arms is negligible with respect to that of the long ones. However, such stress reduction is balanced by a dilution effect whereby the unstretched arms reduce SORF as they decrease the Kuhn-segment order parameter of the system. As a result of that dilution, the stress contributed by the backbone is larger. The two effects compensate one another, so that the overall stress is virtually the same of the other architectures.

Keywords

Viscoelasticity, polymer dynamics, molecular simulations, nonlinear rheology

INTRODUCTION

Some years ago, Ianniruberto and Marrucci[1] proposed a simple molecular picture for entangled branched polymers in fast elongational flows. They argued that branched polymers behave like linear ones because the arms are sucked into the backbone tube, and hence the entire branched molecule becomes effectively linear. Based on this picture, they calculated the effective Rouse time for a few branched architectures (on which data were available[2]) and demonstrated that in fast flows the elongational rate dependence of the steady-state stress is universal if the stress is plotted against the Weissenberg number based on such effective Rouse time.

New data were then provided by Huang et al.[3], who measured the stress in startup experiments of fast uniaxial elongational flows for linear and star polystyrene (PS) entangled melts, with the same span molar mass (i.e., that of the segment, called backbone, with the largest end-to-end distance). In addition to a symmetric three-arm star polymer, they also examined an asymmetric one with a short arm attached to the center of the backbone. They showed that the steady-state stress in fast flows is the same for all three structures, and they interpreted such result as a confirmation of the theory by Ianniruberto and Marrucci.[1]

However, as we shall see, the behavior of these stars is not trivial. For example, in the asymmetric case, if the short arm stays unstretched, its contribution to the total stress is negligible, but at the same time the behavior of the backbone needs to be revisited because the unstretched arm also affects the overall Kuhn-segment order parameter, and as a consequence the monomeric friction reduction (SORF)[4–7] is affected too. As the order parameter is reduced, the friction coefficient stays somewhat larger than in the case where the whole molecule stretches out. Hence, the stress contributed by the backbone is larger. Because the two effects (unstretched arm and larger friction) contribute in an opposite way to the overall stress, one of the objectives of the simulations is finding out which of the two wins. In the symmetric case, as we shall see, the observed behavior also shows some challenging features.

Another issue worth discussing is the transient behavior. Huang et al.[3] show that the stress growth function is virtually insensitive to the branched structure not only at the steady state but also during the startup. This experimental finding is not included in the theory of Ianniruberto and Marrucci[1], who only considered the steady state. The transient stress reflects the development of the orientation and stretch dynamics of the polymers, which, intuitively, would be expected to depend on the branched structure. On such development, theories and/or simulations of fast uniaxial elongational flows appear to be lacking.

In this paper we report on multi-chain sliplink simulations (also known as primitive chain network simulations[8–10]) of linear and star polymers in fast uniaxial elongational flows, mimicking the experiments of Huang et al.[3] Simulations also provide detailed information on stretch and orientation of polymer backbone and short arm separately, and on the effect of the short arm on friction reduction. The results reveal that star polymers behave like linear ones even during the transient. To interpret these results, we also develop new theoretical considerations that augment the suggestion of Ianniruberto and Marrucci[1].

MODEL AND SIMULATIONS

The code used in this study is essentially the same of that previously employed for the simulation of entangled polymers in uniaxial elongational flows.[4,6,11–13] Entangled polymers are replaced by a network, which consists of nodes, strands, and dangling ends. For the case of linear polymers, each polymer chain corresponds to a path connecting two dangling ends through some network nodes and strands. At each node, a sliplink bundles two chains, yet allowing the chains to slide across it. In the case of star polymers, linear segments of chains are connected at a branchpoint. Also branchpoints are network nodes, but chain sliding is forbidden there (crosslinks).

The state variables of the system are the position $\{\mathbf{R}\}$ of the network nodes, the number $\{n\}$ of Kuhn segments on each strand and dangling end, and the number $\{Z\}$ of segments per chain, including strands and dangling ends. Positions $\{\mathbf{R}\}$ obey a Langevin equation describing the force balance among drag force, tension in the connected strands, osmotic force, and random force. The finite-extensibility (FENE) effect on strand tension is accounted for via the FENE-P approximation[11]. Kuhn segment numbers $\{n\}$ are calculated by a Langevin-type kinetic equation, in which a similar 1D tension balance is considered along the chain. As regards the time development of {Z}, we monitor *n* for each dangling end. If *n* becomes lower than a certain threshold, we remove the sliplink to which the dangling segment is connected. Vice versa, if *n* exceeds an upper threshold, we create a new sliplink on the end segment by hooking another segment that is randomly chosen among the surrounding ones.

The above simulation algorithm fairly reproduces the well-known molecular mechanisms of the dynamics of entangled polymers, such as reptation[14], contour length fluctuations[15], and thermal[16–19] and convective[20,21] constraint release. For the case of branched polymers, arm retraction[22] generating a significant constraint release (or tube dilation)[23,24] is automatically accounted for by the same algorithm. However, the reptative motion of the branchpoint following hierarchical relaxation[13,25–30] must be added to the classical algorithm. This is implemented through the following recipe. When the number of sliplinks on a branching arm becomes zero, we allow the sliplink next to the branchpoint to hop across the branchpoint. We also add the branchpoint withdrawal[13,28,31] when the tension in one of the strands emanating from the branchpoint becomes larger than the sum of the tensions in the other strands. The latter mechanism rarely occurs in star polymers.

The SORF mechanism[4–7] is accounted for through the following equations:

$$\frac{\zeta(F_{SO})}{\zeta_0} = \frac{1}{(1+\beta)^\gamma}\left[\beta + \frac{1}{2}\{1 - \tanh[\alpha(F_{SO} - F_{SO}^*)]\}\right]^\gamma \quad (1)$$

$$F_{SO} = \tilde{\lambda}^2 S \quad (2)$$

Here, $\zeta$ is the friction coefficient of the Kuhn segment, and $\zeta_0$ is its equilibrium value. The parameters $\beta$, $\gamma$, $\alpha$, $F_{SO}^*$ are phenomenological. They are set to: $\beta = 5 \times 10^{-9}$, $\gamma = 0.15$, $\alpha = 20$, $F_{SO}^* = 0.14$. $F_{SO}$ is the order parameter of the Kuhn segments due to chain stretch and orientation, $\tilde{\lambda}$ being the stretch ratio normalized to the fully extended conformation, and $S$ the average strand orientation.

The simulations were performed in nondimensional variables, with units of length, energy, and time taken as average strand length $a$ at equilibrium, thermal energy $k_BT$, and strand relaxation time $\tau_0 = n_0\zeta_0 a^2/(6k_BT)$ (with $n_0$ the equilibrium average number of Kuhn segments in each strand), respectively. In the case of PS melts, from the previous studies[4,6,11–13] we take $n_0 = 16$, which corresponds to the strand molar mass $M_0 = M/Z_0 = 11\text{k}$, $Z_0$ being the equilibrium average number of strands per molecule. The examined PS architectures are summarized in Table I. Note that, to the samples of Huang et al.[3], we add Star40, for which the short arm molar mass falls in between Star90 and Star20. From $M_0$, the modulus $G_0$ is obtained as $G_0 = \rho RT/M_0 = 0.29\,\text{MPa}$ at the temperature of the experiments. Note that $M_0$ and $G_0$ are similar to, but different from, the entanglement molar mass $M_e$ and plateau modulus $G_N^0$ usually reported in the literature, as discussed earlier[32,33].

Table I: Examined PS melts

| Sample | Architecture | Long arm | | Short arm | | Total | | $\tau_R(s)$ |
|---|---|---|---|---|---|---|---|---|
| | | $M_w$ | $Z_0$ | $M_w$ | $Z_0$ | $M_w$ | $Z_0$ | |
| Lin180 | Linear | | | | | 187k | 18 | 190 |
| Star90 | 3-arm symmetric star | 92.4k | 9 | 92.4k | 9 | 289k | 27 | 264 |
| Star20 | 3-arm asymmetric star | 92.4k | 9 | 20.5k | 2 | 208k | 20 | 194 |
| Star40 | 3-arm asymmetric star | - | 9 | - | 4 | - | 22 | 205 |

Using these parameters, we run equilibrium simulations and obtained the linear relaxation modulus through the Green-Kubo formula. Eight independent simulation runs were performed, with periodic boundary conditions, and a cubic simulation box of dimension $8^3$. Comparison of the predicted linear viscoelasticity with the experimental data, as shown later, gives $\tau_0 = 2.9$ s. The Rouse time of each architecture is obtained from this $\tau_0$ value (see later), and is reported in Table I. Once known $\tau_0$, we could perform simulations of uniaxial elongation flows at several extensional rates, corresponding to the experimental ones. The elongational simulations started from a thin box of 4x45x45, where 4 is the dimension in the stretching direction. Simulations were run up to when the box elongated close to 500x4x4. The maximum elongation corresponds to a stretch ratio of 125, and to a Hencky strain of 4.8. The adopted simulation method for elongational flows was validated earlier[11].

RESULTS AND DISCUSSION

Figure 1 shows the predicted linear viscoelasticity (LVE) of linear and star melts compared to the PS experimental data. As shown previously[32,34], simulations reasonably reproduce the experiments for the linear polymer. However, for star polymers, simulations predict a somewhat faster relaxation than the experiments. We conjecture that this discrepancy is due to the molar mass distribution. Indeed, even though the experimentally investigated samples are nearly monodisperse, the viscoelasticity of branched polymers is quite sensitive to the inclusion of a few longer chains, as reported earlier[35,36]. The other issue we note is that the plateau modulus of Star20 is considerably lower than that of the other polymers, probably due to the dilution effect of the short arm. This reduction of $G'$ in the plateau region is not correctly captured by our simulations, possibly because of the employed coarse-graining. Indeed, the LVE data were nicely reproduced by multi-chain slip-spring simulations[37], in which the motion of the Rouse segments between entanglements is also accounted for. In any event, the coarse-grained simulations used here reasonably describe the linear viscoelasticity of the symmetric Star90, as well as that of other asymmetric star polymers, for which the short arm is reasonably well entangled[27].

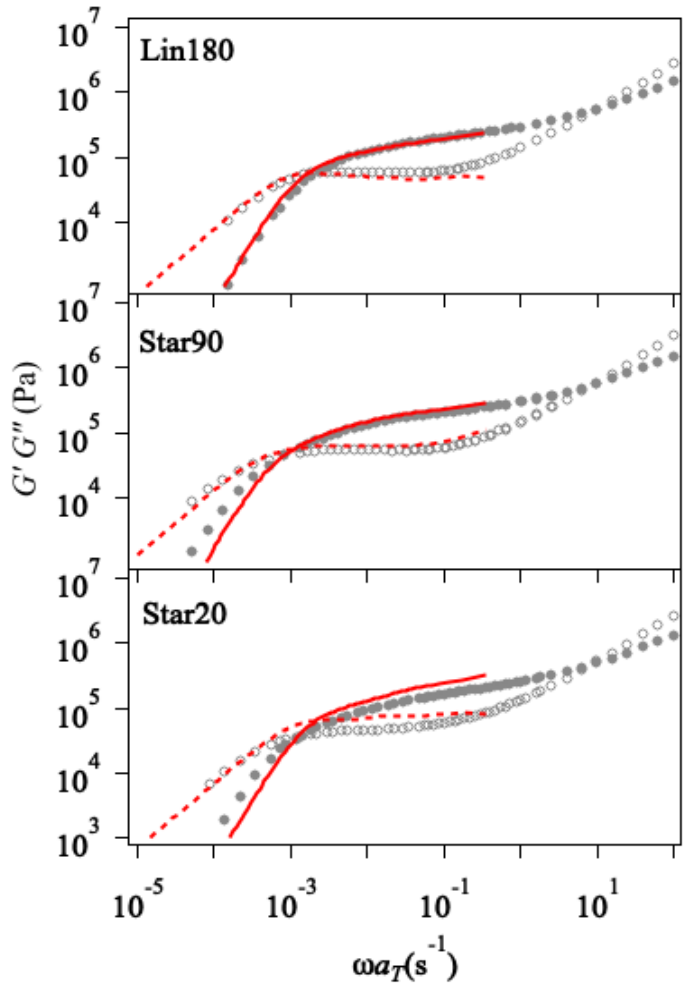

**Figure 1** Linear viscoelasticity of Lin180, Star90 and Star20 at 130°C, from top to bottom. Symbols and lines are experimental[3)] and simulation results, respectively.

Figure 2 shows the viscosity growth during startup of uniaxial elongational flows. The simulations with SORF semi-quantitatively reproduce the experimental data, also for star polymers. As reported earlier[4,6,7)], for PS melts SORF plays a key role in reducing the elongational stress in fast flows. Simulations without SORF significantly overestimate the stress, as shown by the red broken curves in Figure 2. It is fair to note that the linear viscoelastic envelope (black broken curves) is somewhat overestimated for Star20 due to the discrepancy in the plateau region of $G'$ shown in Fig. 1, and previously discussed. Nevertheless, we emphasize that, to our knowledge, this study is the first systematic attempt to reproduce the elongational data of star polymers theoretically (see below), and computationally.

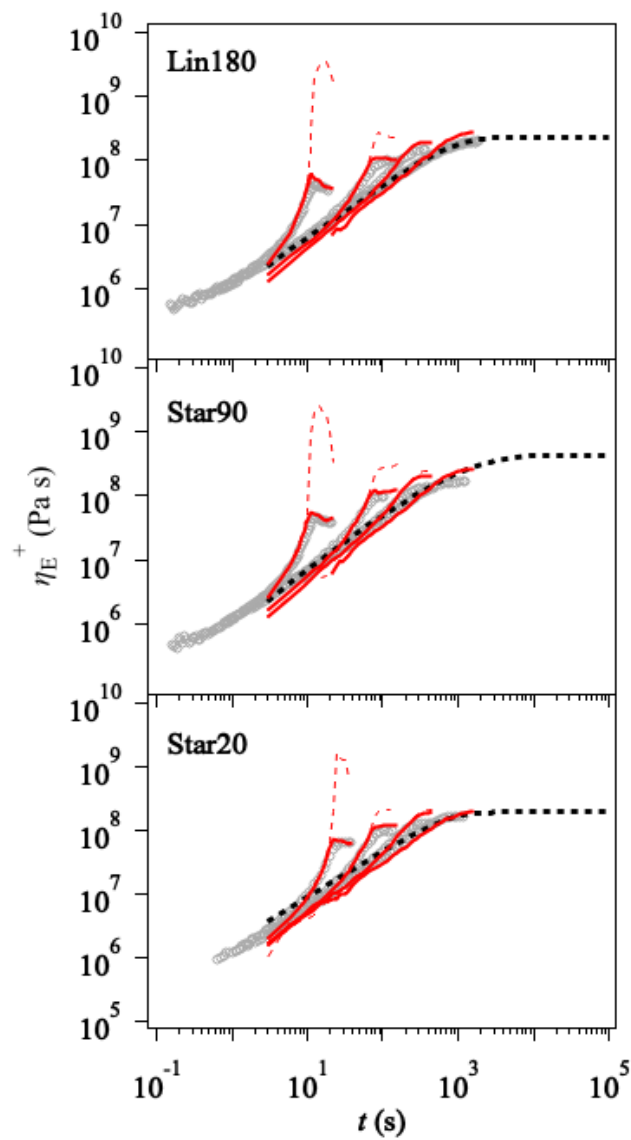

**Figure 2** Viscosity growth during startup of uniaxial elongational flows at 130°C for Lin180, Star90, and Star20, from top to bottom. Elongational rates are 0.2, 0.03, 0.01, and 0.003 s$^{-1}$ for Lin180 and Star 90, whereas they are 0.1, 0.03, 0.01, and 0.003 s$^{-1}$ for Star20, from left to right. Experimental data[3] and simulation results are indicated by gray circles and red curves, respectively. Simulation results without SORF (yet with FENE) are shown by red broken curves. Black broken curves are the predicted linear viscoelastic envelopes.

Figure 3 shows the steady-state viscosity plotted against strain rate. As seen in Figure 2, the simulations semi-quantitatively reproduce the experimental data, and the viscosity is nearly the same for all examined architectures, especially in fast flows. At high rates, the viscosity exhibits a power-law decrease with increasing strain rate, with a power-law exponent of ca. -0.5. This decrease of the viscosity is typical of well-entangled PS melts[38].

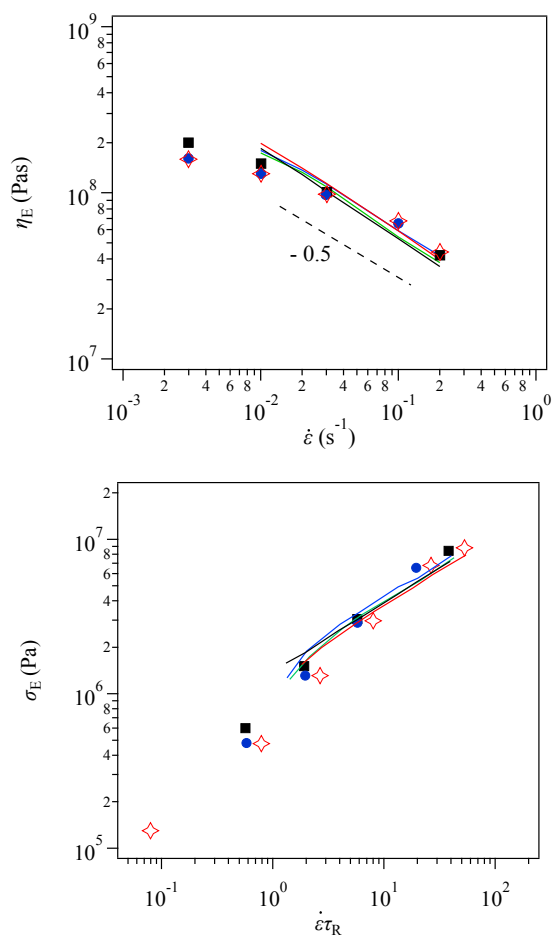

**Figure 3** Steady-state elongational viscosity plotted against strain rate (top) and steady-state stress vs. Rouse-time-based (from values in Table I) Weissenberg number (bottom). Squares, stars, circles are experimental data of Huang et al.[3] for Lin180, Star90, Star20, respectively. Solid curves show simulation results (with SORF) for Lin180 (black), Star90 (red), Star 20 (blue), and Star 40 (green). The broken line in the top panel indicates a slope of -0.5.

As mentioned in the Introduction, Ianniruberto and Marrucci[1] proposed a theory to calculate the Rouse time of entangled branched polymers in fast elongational flows, and they suggested a universal, molar-mass and architecture independent flow curve in fast flows, if the elongational stress is plotted against the Rouse-time-based Weissenberg number. In order to test their suggestion with all architectures here considered (and not examined by Ianniruberto and Marrucci[1])), we need to calculate the Rouse times $\tau_R$ for the symmetric and asymmetric 3-arm star polymers, defined as the reciprocal stretch rate above which chains get stretched by the flow[1].

Let us call $\tau_{R,b}$ the Rouse time of the backbone. When $Wi_{R,b} = \dot{\varepsilon}\tau_{R,b} \approx 1$, the backbone tube is certainly fully oriented in the stretching direction. Hence, putting the origin of Cartesian coordinates at the branchpoint, and the $x$ axis along the backbone, the positions of the backbone ends are $-0.5$ and $+0.5$, respectively, if $x$ is made nondimensional as ratio to the length $L$ of the backbone ($L$ obviously being twice the length of the long arms). Now, let's call $\tau_a$ the orientational relaxation time of the short arm. Since $Wi_{R,b} = \dot{\varepsilon}\tau_{R,b} \approx 1$, the short arm is either unoriented or oriented parallel to the backbone depending on whether it is $\tau_a < \tau_{R,b}$ or $\tau_a > \tau_{R,b}$, respectively. In the former case, the friction center is obviously at the branchpoint ($x = 0$), and the Rouse time $\tau_R$ of the asymmetric 3-arm star coincides with $\tau_{R,b}$. Conversely, the case $\tau_a > \tau_{R,b}$ implies that the friction center is off from the branchpoint, and some calculations are therefore required.

Without loss of generality, we may assume that the short arm orients itself in the positive $x$ direction, and let us call $d > 0$ the unknown nondimensional position of the friction center. Further, call $r = L_a/L$ the ratio of the length $L_a$ of the short arm to the length $L$ of the molecule backbone ($0 < r \leq 0.5$), where $r = 0.5$ corresponds to the symmetric star. The tension at the friction center is then calculated from the contributions $F_r$ and $F_l$ to the right and to the left of it, respectively,

$$F_r = \int_d^{0.5} \hat{\zeta}\dot{\varepsilon}L^2(x-d)dx + \int_d^r \hat{\zeta}\dot{\varepsilon}L^2(x-d)dx = \hat{\zeta}\dot{\varepsilon}L^2\left[d^2 - \left(r+\frac{1}{2}\right)d + \frac{1}{2}\left(r^2+\frac{1}{4}\right)\right],$$

$$F_l = \int_{-0.5}^d \hat{\zeta}\dot{\varepsilon}L^2(d-x)dx + \int_0^d \hat{\zeta}\dot{\varepsilon}L^2(d-x)dx = \hat{\zeta}\dot{\varepsilon}L^2\left(d^2 + \frac{1}{2}d + \frac{1}{8}\right), \tag{3}$$

where $\hat{\zeta}$ is the friction coefficient of the molecule per unit length of the confining tube, and $\dot{\varepsilon}$ is the rate of the extensional flow. Since $F_r$ and $F_l$ must be equal to one another, we obtain

$$d = \frac{1}{2}\frac{r^2}{1+r}, \tag{4}$$

and for the tension $F = F_r = F_l$ at the friction center we can simply use the second Eq.

(3):

$$F = \frac{\hat{\zeta}\dot{\varepsilon}L^2}{8}(1 + 4d + 8d^2) \ , \tag{5}$$

By progressively increasing $\dot{\varepsilon}$, chain stretch will begin when the friction tension $F$ becomes comparable to the equilibrium tension in the tube, i.e., to $3\,k_BT/a$. There follows that the Rouse time $\tau_R$ of the whole three-arm star polymer is given by

$$\frac{\tau_R}{\tau_{R,b}} = 1 + 4d + 8d^2 \ , \tag{6}$$

where we adopted for $\tau_{R,b}$ the expression $\tau_{R,b} = \hat{\zeta}L^2 a/(24 k_B T)$, to which $\tau_R$ obviously reduces when $d = 0$, i.e., when the length of the short arm vanishes. Notice, however, that the correct expression for $\tau_{R,b}$ should be $\tau_{R,b} = \hat{\zeta}L^2 a/(3\pi^2 k_B T)$ (see e.g. Doi and Edwards[39]). The minor difference in the numerical factor is due to the fact that the above calculation is correct to within the order of magnitude.

Now, by recalling that $\tau_0 = n_0\zeta_0 a^2/(6k_BT) = \hat{\zeta}a^3/(6k_BT)$, the exact expression for $\tau_{R,b}$[39] is then linked to $\tau_0$ through the ratio $\tau_{R,b}/\tau_0 = 2L^2/(\pi^2 a^2) = 2Z_0^2/\pi^2$. By also recalling that $\tau_0 = 2.9$ s, and that for the linear polymer it is $Z_0 = 18$, we get $\tau_{R,b} = 190$ s. Moving on to the star molecules, we have $r = 0.5, 0.22, 0,11$ for Star90, Star40, Star20, respectively. The corresponding $d$-values are obtained from Eq. (4), and finally the Rouse times are calculated from Eq. (6) and listed in Table I. It should, however, be noted that we have tacitly assumed that for the two asymmetric star polymers the short arm aligns to the elongation direction. In any event, Table I shows that differences in the Rouse time between the short-arm architectures and the linear polymer are actually minor.

An important remark concerns the 3-arm symmetric star. Indeed, the Rouse time of Star90 comes out larger than that of Lin180 by ca. 40%. As shown above, the reason for such difference is the displacement of the friction center from the branchpoint. Such difference would imply that, at the same (large) strain rate, the elongational viscosity of Star90 is larger than that of Lin180. The upper panel of Fig. 3, however, shows that those two viscosities are in fact undistinguishable in fast flows, probably because a 40% difference in the abscissa values is hardly visible in Fig. 3. For the same reason, the bottom panel in Figure 3, reporting the steady state elongational stress vs. the Weissenberg number based on the Rouse times of Table I, shows that Star90 and Lin180 fall on the universal flow curve predicted by Ianniruberto and Marrucci[1], together with Star20.

A further possible reason for the indistinguishability between Star 90 and Lin180 can be found in flow-induced intermolecular interactions that bring the friction center closer to the branchpoint. Such is the case, for example, of the molecule pair conformation schematically depicted in Figure 4, which may form during the relative motion between fully aligned molecules in fast elongational flows. In Figure 4, it is also reported a snapshot of one such pair observed in the simulation. It must be said, however, that in our Brownian simulations the event depicted in Figure 4 is expected to be highly improbable because strands are phantom, and topological interactions only take place via sliplinks. Molecular dynamics simulations would be required to confirm the existence of such aggregates in large extent. On the other hand, the paired conformation schematically depicted in Figure 4, as well as higher aggregates, are expected to be stable in fast elongational flows, as is the case of ring polymer pairs reported in the molecular dynamics simulations of O'Connor et al[40].

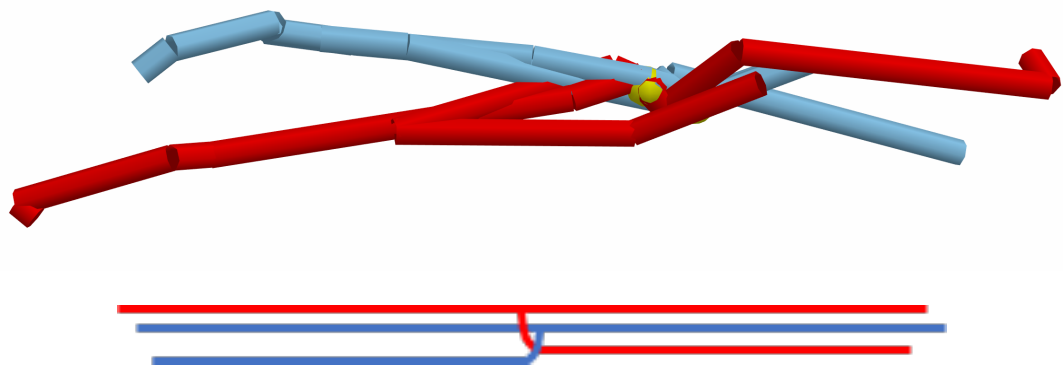

**Figure 4** Top: A snapshot of a pair of hooked 3-arm-star molecules in fast elongational flow ($\dot{\varepsilon} = 0.03 \text{ s}^{-1}$). The yellow dot indicates the common position of the two branchpoints. Bottom: Schematic representation of the aggregate.

Let us now discuss the behavior of the asymmetric stars in better detail. To such purpose, it is convenient to use the decoupling approximation for the stress, mentioned in the Introduction. Figure 5 shows, from top to bottom, the elongational stress, the stretch $\lambda$, the tube segment orientation $S$, the entanglement number $Z$, and the friction coefficient $\zeta$. In the upper panel, the total stress essentially coincides with the contribution of the long arms, while that of the short arms (dashed lines) is definitely minor. This is confirmed by the intermediate panels, showing that both the stretch and the orientation of the short arms are much lower than those of the long ones.

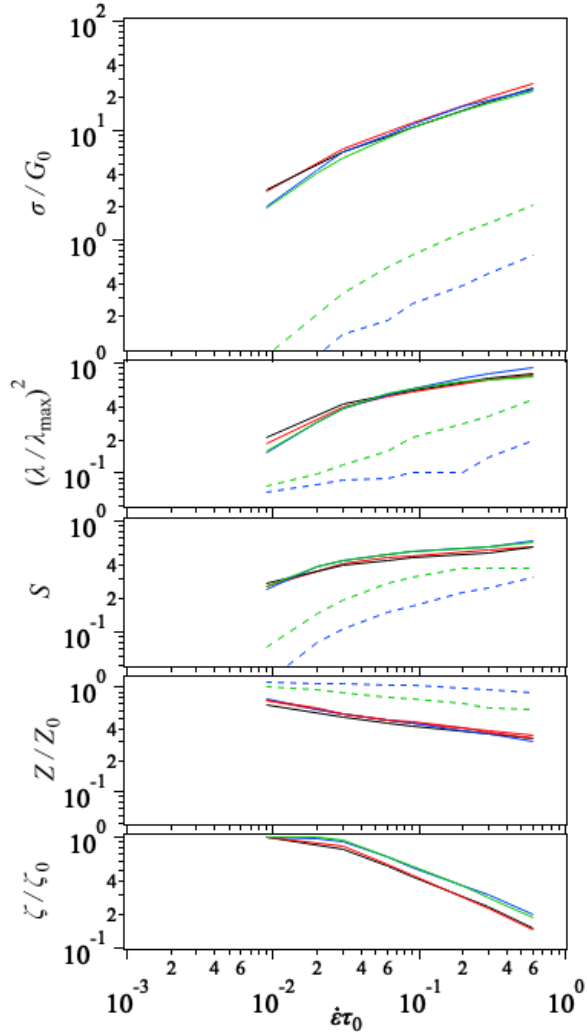

**Figure 5** From top to bottom, steady-state uniaxial stress, stretch, orientation, entanglement number, and friction coefficient (all suitably normalized) vs. $\dot{\varepsilon}\tau_0$. Curves in black, red, blue and green show the results for Lin180, Star90, Star20, and Star40, respectively. For the asymmetric stars, solid and broken curves indicate the results for the long arm and the short one, respectively.

Results in Figure 5 then confirm that the short arms are very weakly affected by the flow, marginally contributing to the stress, and essentially behaving as a diluent for the long arms. The question then arises why the stress of the asymmetric stars coincides with that of the linear polymer in spite of the fact that the backbones are sort of diluted. The solution of this "mystery" can be found in the bottom panel of Figure 5 showing that the friction coefficient of the asymmetric star polymers falls slightly above that of the linear one. The larger friction coefficient is due to the fact that the short arms, being less oriented and stretched, lower the average value of the order parameter. Hence, it so happens that the reduced contribution to the stress of the short arms is essentially compensated by the increased contribution of the long ones due to a larger friction coefficient.

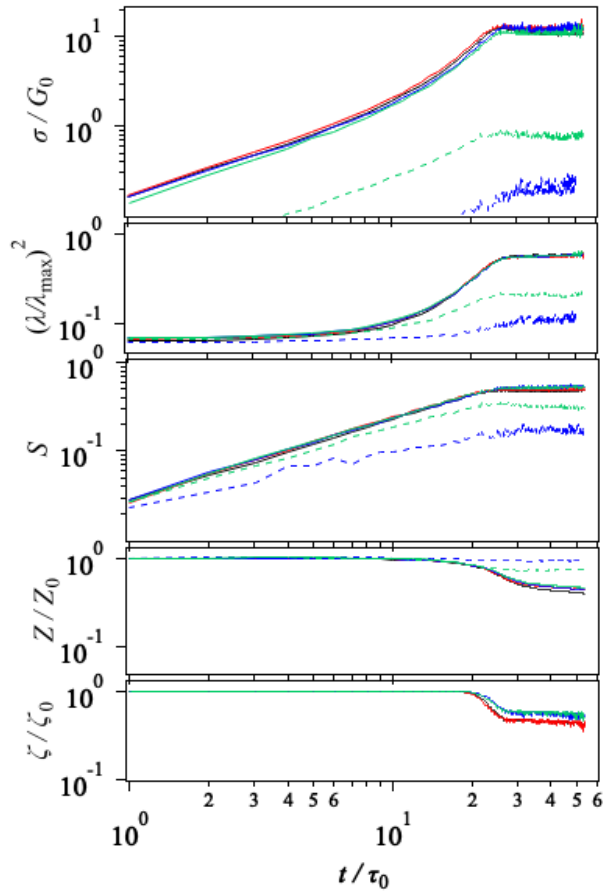

**Figure 6** Transient startup curves for stress, stretch, orientation, entanglement number, and friction coefficient (from top to bottom) at the rate $\dot{\varepsilon} = 0.03 \text{ s}^{-1}$. Curves in black, red, blue, and green are results for Lin180, Star90, Star20, and Star40, respectively. For the asymmetric stars, solid and broken curves refer to long and short arms, respectively.

Figure 6 shows the same quantities of Figure 5 during the transient startup at the rate $\dot{\varepsilon} =$

0.03 s$^{-1}$. The top panel confirms that, also during the transient, the stress contributed by the backbones (solid curves) is insensitive to the branching structure. A similar insensitivity to the molecular structure is also found for the stretch, the orientation, and the entanglement number. The backbones of the star polymers, just as the linear polymers, first orient to the elongation direction (up to $t/\tau_0 \approx 10$), then start stretching, so that the stress steeply grows all the way up to the steady state, a behavior known as strain hardening, (i.e., upwards deviation from the linear viscoelastic envelope). When the stretch approaches the steady state, convective constraint release takes over, and the entanglement density decreases. This chain dynamics is well known for linear polymers, but a similar behavior for the backbone of star polymers is reported here for the first time. It is finally noted that, as expected, the friction coefficient starts decreasing when chains start stretching, but the important information is that such flow-induced reduction is larger in the linear polymers (and in the symmetric star) than in the asymmetric case, as already noted for the steady state results of Figure 5.

CONCLUSIONS

We have performed primitive chain network simulations of fast uniaxial elongational flows for linear and star polystyrene melts, all with the same span molar mass. The simulated viscosity growth curves were in semi-quantitative agreement with the experimental data and were found to be insensitive to the molecular structure, in full agreement with data.

For the symmetric star polymer, such insensitivity is due to the fact that its Rouse time falls close enough to that of the linear polymer, thus masking possible differences. Furthermore, it appears quite possible that flow-induced molecular aggregates form in fast extensional flows (see an example in Figure 4), which have a Rouse time even closer to that of the linear polymer.

In the asymmetric star case, a balancing effect explains the matching with the linear polymer case. On the one hand, short arms marginally contribute to the stress. On the other hand, their dilution effect on the long arms makes the friction coefficient larger than in the linear polymer case, thus increasing the stress contribution of the long arms. This is confirmed by the detailed analysis in Figures 5 and 6.

In conclusion, the underlying theory for fast elongational flows of branched polymers,

first suggested by Ianniruberto and Marrucci[1], is here confirmed and augmented.

ACKNOWLEDGEMENTS

YM wishes to express sincere thanks to Prof Tadao Kotaka for his kind encouragement. YM was financially supported by Ogasawara foundation, JST-CREST (JPMJCR1992), and NEDO (JPNP16010).